\newcommand{\rthpeak}{R_{\rm th}^{\rm peak}}
\newcommand{\rthint}{R_{\rm th}^{\rm int}}
\documentclass[aps,preprint, prl,superscriptaddress, showpacs]{revtex4}
\usepackage{graphicx}
\usepackage{amsmath}
\usepackage{amssymb}
\usepackage{amsthm}
\usepackage{pdfpages}

\begin{document}


\title{Classification of light sources and their interaction with active and passive environments}


\author{Ramy G. S. El-Dardiry}
\email[]{dardiry@amolf.nl} \homepage[]{http://www.randomlasers.com}
\affiliation{FOM-Institute for Atomic and Molecular Physics AMOLF,
Science Park 104, 1098 XG Amsterdam, The Netherlands}

\author{Sanli Faez}
\affiliation{FOM-Institute for Atomic and Molecular Physics AMOLF,
Science Park 104, 1098 XG Amsterdam, The Netherlands}

\author{Ad Lagendijk}
\affiliation{FOM-Institute for Atomic and Molecular Physics AMOLF,
Science Park 104, 1098 XG Amsterdam, The Netherlands}


\date{22 December 2010}

\begin{abstract}
Emission from a molecular light source depends on its optical and
chemical environment. This dependence is different for various
sources. We present a general classification in terms of Constant
Amplitude and Constant Power Sources. Using this classification, we
have described the response to both changes in the LDOS and
stimulated emission. The unforeseen consequences of this
classification are illustrated for photonic studies by random laser
experiments and are in good agreement with our correspondingly
developed theory. Our results require a revision of studies on
sources in complex media.
\end{abstract}

\pacs{42.25.Dd, 42.55.Zz, 32.50.+d}

\maketitle

Atomic and molecular light sources are essential tools in the
natural sciences. Physicists use these light sources in a great
variety of situations, for example to study light-matter
interactions in the context of cavity quantum electrodynamics
\cite{Wilk2007, Sapienza2010}, to probe vacuum fluctuations inside
and around photonic and plasmonic nanostructures
\cite{Koenderink2002, Farahani2005}, or as building blocks for
lasers \cite{Tureci2008}. In the life sciences, fluorescent proteins
have quickly become one of the most important workhorses soon after
their discovery \cite{Tsien1998}. Major engineering efforts are
nowadays devoted to inventing light-source based microscopy
techniques, in order to obtain improved resolution and sensitivity
\cite{Hell2005, Min2009}.

The prominence of light sources in scientific experiments solicits
for a well-defined classification of different types of sources. We
propose such a classification analogous to the field of electronics
where every circuit design incorporates a well defined source. In
electronics, ideal sources are classified as Constant Current
Sources (CCS) or Constant Voltage Sources (CVS) depending on their
response to a certain load \cite{Horowitz}.

Mathematically, a point source (sink) is incorporated by a positive
(negative) divergence ($S=\nabla\cdot {\bf J}$) of a certain vector
quantity in space. In order to be classified as a source for light,
light should either be created by conversion from a different type
of energy, e.g. by electroluminescence, or by a photochemical
process in which the absorbed excitation photon differs in frequency
from the emitted photon, e.g. in three- and four-level systems. In
contrast, two-level systems cannot be considered as light sources,
they are scatterers instead. Although we limit ourselves in this
manuscript to a discussion of four-level systems, our approach is
general and can be applied to other light generation mechanisms as
well.

In a four-level system there are in general two decay channels from
the lowest vibrational sublevel of the excited state to a
vibrational sublevel of the ground state: a radiative and a
nonradiative channel. These two relaxation mechanisms are competing
for the number of molecules in the excited state. In a similar way
as two parallel resistances are competing for current in a simple
electronic CVS circuit. The quantum efficiency of the molecule
describes the ratio between the radiative and total decay rate. To
qualify for a Constant Power Source (CPS) the power emitted by the
radiative channel must be independent on any change in the ``load''
of the radiative decay channel. In a Constant Amplitude Source (CAS)
the number of transitions is conserved, but the power emitted by the
source is dependent on the conductivity of the radiative decay
channel.

In this Letter, we study the influence of light source typology on
the generation of light in complex media. We provide a clear
demonstration of the relevance of our classification with new random
laser experiments, where different kinds of light sources act as
different gain media. Besides these new experimental results we
provide a description of the interaction of light sources with their
environment by calculating the power emitted by a light source in
the vicinity of a single scatterer. Our theoretical and experimental
studies emphasize that the response of a light source to either
stimulated emission or a change in the Local Density of States
(LDOS) depends on its class. In the end we discuss the impact of CPS
and CAS in studies on light sources and multiple scattering.

\begin{figure}
  \centering
  \includegraphics[width=0.4\textwidth]{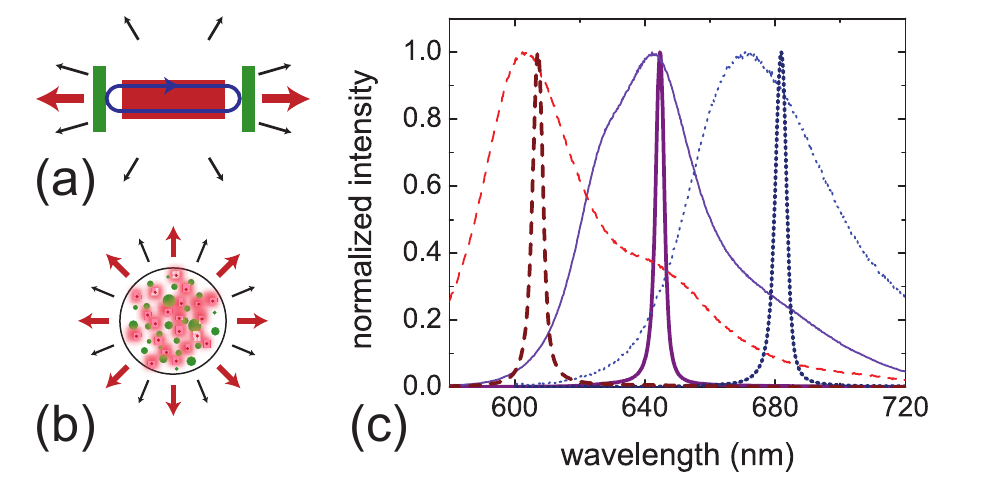}\\
  \caption{Illustration of emission directionality below threshold (black arrows) and above threshold (red arrows)
  in (a) a conventional laser and (b) a random laser.
  In a random laser the emitted light by an ensemble of sources is always omnidirectional.
  (c) Experimental emission spectra below and narrowed spectra above random laser threshold for three different light sources: Rhodamine 640 P
  (red dashed lines), Cresyl Violet (purple solid lines), and Nile Blue (blue dotted
  lines). The $\beta$-factor is determined by the quotient of the
  area of the normalized spectra above and below threshold.
  }\label{gexperiment}
\end{figure}

\textit{Experiment - } In a random laser \cite{Wiersma2008} the role
of sources is twofold: first, they are seeds of spontaneous light
emission; second, they amplify light by stimulated emission of
radiation. Due to the multiple-scattering feedback mechanism, random
lasers form a unique laser system. In contrast to conventional
lasers, they have a statistically isotropic mode selectivity as
illustrated by the cartoons in Fig.~\ref{gexperiment}(a) and (b).
The mode selection is solely determined by the spectral shape of the
gain curve. In a random laser, measuring the emitted energy into a
large enough solid angle corresponds to measuring the total emitted
intensity: diffusion mimics an integrating sphere. In our
experiments, we utilize this much neglected property of random
lasers to study the energy emitted by light sources with different
quantum efficiencies for varying pump rates.

Three molecular light sources were studied in a random laser
configuration by suspending titania particles (R900 DuPont, volume
fraction 1\%) into three different 1 mM solutions of organic dyes in
methanol. The three dye solutions acted as gain media and were
chosen based on their quantum yields ($\phi$) reported in literature
\cite{Siegman, Isak1992}: Rhodamine 640 P ($\phi=$ 1), Cresyl Violet
($\phi=$ 0.54), and Nile Blue ($\phi=$ 0.27), see Methods in the
Supplementary Material \footnote{See EPAPS Document No. [number will
be inserted by publisher] for [give brief description of material].
For more information on EPAPS, see
http://www.aip.org/pubservs/epaps.html.} for more information on the
used optical setup and the sample preparation.

\begin{figure}
  \centering
  \includegraphics[width=0.35\textwidth]{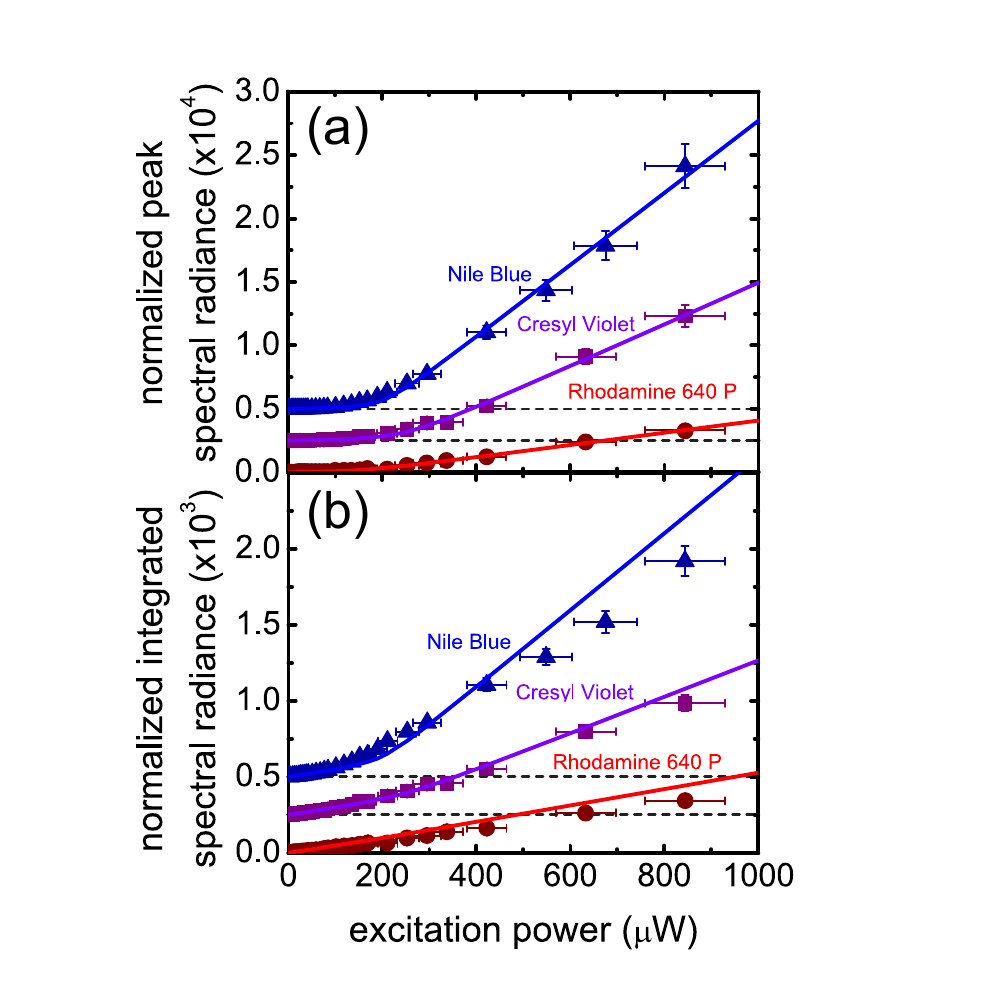}\\
  \caption{Input-output diagrams for random lasers consisting of light sources with low and high quantum efficiencies.
  (a) peak spectral radiance versus pump power for three random lasers with different molecular light sources. The solid lines are
  fits to the experimental data. (b) integrated spectral radiance versus pump power for three random lasers. The solid lines are
  theoretical calculations. The Rhodamine 640 P random laser does not show a clear threshold. All data points in (a) and (b) were normalized
  to the values at 2.1 $\mu$W and the results for the Nile Blue and
  Cresyl Violet random lasers were shifted vertically for clarity.
  }\label{gexp_results}
\end{figure}

For all random laser samples, the fluorescent emission spectra were
recorded for different values of the pump fluence below and above
threshold. In Fig. \ref{gexperiment}(c) emission spectra far below
and far above threshold are plotted. The spectra above threshold are
narrower by a factor $\thicksim$ 10 compared to the spectra below
threshold and the peaks are slightly red shifted due to
reabsorption. Figure \ref{gexp_results} shows (a) the peak and (b)
the integrated spectral radiance versus the excitation power. The
peak spectral radiance shows a clear threshold for all the three
random laser systems. In a conventional laser angular redistribution
of light emission causes a threshold in the spectrally integrated
power of the output beam irrespective of the chosen gain medium.
However, in the experimental results shown in Fig.
\ref{gexp_results}(b) we observe that for the random laser with the
highest quantum efficiency gain medium (Rhodamine) such a threshold
in the integrated spectral radiance is absent.

\textit{Random laser model - } Standard lasers are described with
rate equations \cite{Siegman} describing the number of photons in
the cavity mode, $q(t)$, and the number of molecules in the upper
laser level, $N(t)$. For a four-level system it is usually assumed
that only the population of the ground state and the upper-laser
level are significant. To model our random laser experiment, we
extend such a set of equations with an equation describing the
number of photons, $w(t)$, emitted outside the lasing mode
\begin{align}
    \frac{\textrm{d} q}{\textrm{d} t} &=
    -q\gamma_{\rm c}+\beta\gamma_{\rm r}Nq+\beta\gamma_{\rm r}N,\label{rate_eqn_photons2}\\
    \frac{\textrm{d} w}{\textrm{d} t} &=
    -w\gamma_{\rm c}+N\gamma_{\rm r}(1-\beta),\label{rate_eqn_photonswing}\\
    \frac{\textrm{d} N}{\textrm{d} t} &=
    R-N\gamma_{\rm tot}-\beta\gamma_{\rm r}Nq\label{rate_eqn_photonsupperlaser}.
\end{align}
Here, $R$ is the pump photon rate, $\gamma_{\rm c}$ is the cavity
decay rate and $\gamma_{\rm tot}$ is the total decay rate with
$\gamma_{\rm tot} = \gamma_{\rm r}+\gamma_{\rm nr}$ where
$\gamma_{\rm r}$ and $\gamma_{\rm nr}$ are the radiative and
nonradiative decay rates respectively. The spontaneous emission
factor $\beta$ describes what fraction of the spectrum contributes
to the lasing emission \cite{GvS2002}. Due to the absence of angular
mode selection in a random laser, the $\beta$-factor suffices for
distinguishing photons inside and outside the lasing mode: for
photons emitted in the wings of the spectrum stimulated emission is
neglected in rate equation (\ref{rate_eqn_photonswing}), whereas for
photons emitted into the peak of the spectrum, Eq.
(\ref{rate_eqn_photons2}), stimulated emission is added to the
spontaneous emission rate. The random lasers considered here have a
smooth spectrum above threshold. The particular case of a random
laser with spectral spikes \cite{Cao1999} requires a different
formulation \cite{Tureci2008}. We determine the $\beta$-factor for
the three random lasers by calculating the ratio of the integrated
spectra above and below threshold after normalizing to the peak
value \cite{GvS2002}: for Rhodamine $\beta=0.099$, for Cresyl Violet
$\beta=0.088$, and for Nile Blue $\beta=0.076$.

To infer the threshold, the steady-state solutions to Eqs.
(\ref{rate_eqn_photons2}-\ref{rate_eqn_photonsupperlaser}) for the
number of photons in the peak and the wings of the spectrum are
calculated
\begin{align}
    q&=-\frac{1}{2\beta\phi}+\frac{R}{2\gamma_c}+\frac{1}{2}\sqrt{\left(\frac{1}{\beta\phi}-\frac{R}{\gamma_c}\right)^2+4\frac{R}{\gamma_c}}\label{solution_q},\\
    w&=\left(\frac{R}{\gamma_c}-q\right)\frac{1-\beta}{\phi^{-1}-\beta}\label{solution_u}. 
\end{align}
Above threshold the slope of the solution for $q$ changes and the
$\beta$-factor and $\phi$ determine the ``smoothness'' of the
transition. Obtained analytical expressions for the threshold for
the peak and integrated spectral radiance are
\begin{align}
    \rthpeak &= \left[(\beta \phi)^{-1}-1\right]\gamma_{\rm c}\label{solution_peak},\\
    \rthint &= \left[(\beta \phi)^{-1}-{\beta^{-1}}\right]\gamma_{\rm
c}\label{solution_int}. 
\end{align}
Thus, it is wrong practice to use $\rthint$ to find the threshold of
a random laser, because $\rthint \rightarrow 0$ when $\phi
\rightarrow 1$.

A fit of the experimental peak spectral radiance with Eq.
(\ref{solution_q}) gives $\phi$ and $R/\gamma_c$. This second fit
parameter scales the power axis. These fits are shown in Fig.
\ref{gexp_results}(a) and yielded the following values for the
quantum efficiency: Rhodamine $\phi=0.88\pm0.11$, Cresyl Violet
$\phi=0.39\pm0.07$, and Nile Blue $\phi=0.19\pm0.03$. A systematic
deviation might be caused by the method used for estimating the
$\beta$-factor\cite{GvS2002}. A single random laser experiment thus
suffices for analyzing the quantum efficiency of a light source in a
complex medium. The fitted values for $\phi$ are systematically
lower than the their literature values, which we attribute to the
relatively high concentrations of dye molecules in our
experiments\cite{Isak1992}. Using Eq. (\ref{solution_u}) and the
measured values for $\beta$ and $\phi$ we can make a theoretical
prediction for the integrated spectral radiance versus excitation
power. These theoretical curves are plotted in Fig.
\ref{gexp_results}(b) and are in great agreement with the
experimental data.

The remarkable observation of a different behavior of the integrated
spectral radiance, that is the total emitted power, for the three
random lasers as a function of input power is well explained by the
concept of CPS and CAS. The random laser threshold indicates the
transition from spontaneous emission to stimulated emission as the
main mechanism of radiation. In the case of a gain medium consisting
of sources with near unity quantum efficiency, this transition does
not influence the ratio between the number of excitation photons
that are absorbed and the number of photons that are emitted: the
total emitted power scales linearly with the total absorbed power.
Hence, we classify these high quantum efficiency dye molecules as
CPS for light. The threshold in the peak spectral radiance simply
indicates the energy is spectrally redistributed from the wings to
the peak of the spectrum. For a gain medium consisting of sources
with a low quantum efficiency, the transition from spontaneous
emission to stimulated emission also changes the ratio between the
non-radiative and the radiative decay channel. The number of
transitions is conserved but the load of the radiative decay channel
is decreased causing the total emitted power to scale non-linearly
with the pump power. Low quantum efficiency molecules should be
classified as CAS for light as will be discussed in the following
paragraphs.

\begin{figure}
  \centering
  \includegraphics[width=0.35\textwidth]{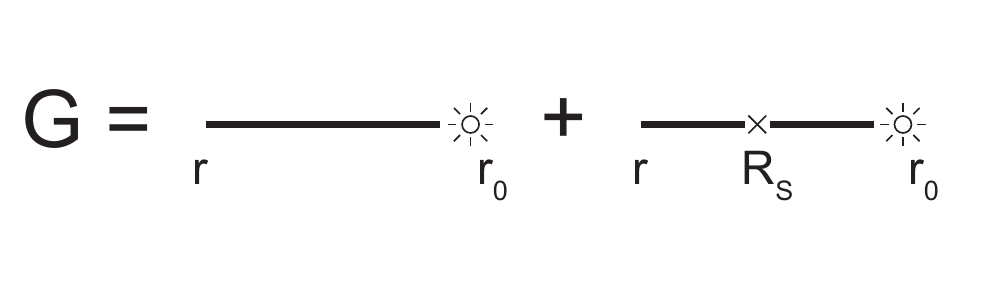}\\
  \caption{Diagram of a Green function describing propagation from a constant amplitude source at ${\bf r_0}$ to ${\bf r}$ with one possible scattering event at
  $\bf{R_{\rm s}}$.
  }\label{gsourceandscatterer}
\end{figure}

\textit{A single scatterer and a source - } We just showed how
invoking stimulated emission of radiation changes the ``resistance''
of an optical transition, alternatively one can change the Local
Density of States (LDOS) at the position of the source. Let us start
by analyzing the output power of a widely used classical dipole
source and then introduce a generalized expression for a source
based on a rate equation analysis. For a point source,
\begin{equation}\label{eqSource}
    j({\bf r},t)=j_0\delta({\bf r-r_0})\textrm{exp}(-i\omega t) +
\textrm{c.c.}
\end{equation}
the output power is related to the LDOS. Figure
\ref{gsourceandscatterer} is a schematic representation of the Green
function, $G$, describing propagation to ${\bf r}$ from a unit
source ($j_0=1$) located at ${\bf r_0}$ in presence of a scatterer
at ${\bf R_s}$. In a homogeneous background, this Green function is
given by
\begin{equation}\label{Greensfunctionsourcescat}
    G_\omega({\bf r, r_0}) = G_\omega^0({\bf r - r_0})+G_\omega^0({\bf r - R_s})t(\omega)G_\omega^0({\bf R_s -
    r_0}).
\end{equation}
Here $G_\omega^0$ is the free space Green function and $t(\omega)$
is the $t$-matrix of the scatterer. To find out the power, $P_{\rm
src}$, radiated by the source in Eq. (\ref{eqSource}) we integrate
the divergence of the current for an infinitesimally small volume
around the source and find%
\begin{equation}\label{Powerexpr}
    P_{\rm src}^{\rm CAS}/P_0=-\frac{4\pi c}{\omega}\textrm{Im}G_\omega({\bf r_0,r_0})\equiv \frac{4\pi^2 c^3}{\omega^2}
    \textrm{LDOS}({\bf r_0}, \omega),
\end{equation}
where $P_0$ is the emitted power without the scatterer present and
the final step is only valid for absorption-free environments. We
prefer to phrase our discussion in terms of LDOS, but the reader is
notified that external absorption can easily be included by
replacing the LDOS with $-\frac{\omega}{\pi
c^2}\textrm{Im}G_\omega$.

The emitted power is thus dependent on the LDOS, which acts as the
inverse of a load on the source. Since the emitted power can both be
higher and lower compared to the vacuum situation, the source we
introduced in Eq. (\ref{eqSource}) is clearly not a CPS; rather we
classify it as a CAS originating from the constant amplitude in Eq.
(\ref{eqSource}). However, from a steady-state rate equation
analysis, explained in full for the interested reader in the
Supplementary Material, we derive that the photon production rate
for a source with nonradiative and radiative decay channels is
proportional to $\gamma_{\rm r}\frac{\gamma_{\rm e}}{\gamma_{\rm
r}+\gamma_{\rm nr}}$, where $\gamma_{\rm e}$ is the excitation rate.
We assume the excitation rate to be constant and independent from
the environment. Our analysis can easily be extended, however, for
environment dependent excitation rates. Therefore the emitted power
reads
\begin{equation}\label{Powerexpr3}
    P_{\rm src}/P_0=\frac{\gamma_{\rm r}}{\gamma_{\rm r}+\gamma_{\rm
    nr}}/\frac{\gamma_{\rm r}^{\rm (0)}}{\gamma_{\rm r}^{\rm (0)}+\gamma_{\rm nr}}=\frac{4\pi^2 c^3}{\omega^2}{\rm LDOS}({\bf r_0}, \omega)\frac{\gamma_{\rm r}^{\rm (0)}+\gamma_{\rm nr}}{\gamma_{\rm r}+\gamma_{\rm
    nr}}.
\end{equation}
Where in the final expression we have replaced the radiative decay
rate with the LDOS using Fermi's golden rule that states that in
nonabsorbing media $\gamma_{\rm r}=A\times{\rm LDOS}$ with $A$
defined as an atomic factor. From this equation it becomes clear
that Eq. (\ref{eqSource}) should be adjusted in a similar way
\begin{equation}\label{Source}
    j({\bf r},t)=\sqrt{\frac{\gamma_{\rm e}}{\gamma_{\rm r}+\gamma_{\rm nr}}}\delta({\bf r-r_0})\textrm{exp}(-i\omega t) +
    \textrm{c.c.}.
\end{equation}
In deriving this expression, we implicitly make the generally valid
assumption that atomic coherence decays very fast. From Eq.
(\ref{Source}) we deduce that the axiomatic expression
(\ref{eqSource}) only applies to a four-level source when
$\gamma_{\rm nr}\gg\gamma_{\rm r}$, a situation often avoided in
experiments. If this condition is not fulfilled, for instance in the
case of a CPS, the strength of the source depends explicitly on the
radiative decay rate and therefore the LDOS. This dependence of
power on the environment is valid for any complex system.

To find the correct wave function from a single source or collection
of sources,
\begin{equation}\label{eqwavefunction}
    \psi(\textbf{r}) = \int G_\omega({\bf r,r_0}) j \{G_\omega({\bf r_0,r_0})\}{\rm
d}{\bf r_0},
\end{equation}
then becomes very involved since it requires knowledge of the Green
function for both the propagation and the generation of light.
Although this dramatically hinders analytic calculations, it should
be straightforward to correctly adjust the source strength in
numerical calculations. Introducing stimulated emission into our
analysis and eventually into Eq. (\ref{Source}) leads to an increase
of the radiative decay rate. This increase leaves a CPS unaltered,
but a CAS will start to look more like a CPS. Stimulated emission
and the LDOS can thus be used to engineer light sources with
$\gamma_{\rm r}/\gamma_{\rm nr}$ as control parameter.

\textit{Conclusion and discussion - } We have developed a new
classification scheme for light sources. Sources with unit quantum
efficiency are classified as constant power sources for light and
those with a low quantum efficiency are classified as constant
amplitude sources. We demonstrated that this classification directly
influences the interpretation of photonic experiments. In the case
of a CAS, both stimulated emission and changes in the LDOS alter the
load of the radiative transition and thereby the output power.

Our classification of light sources is applicable to all photonic
systems. In random media, recently predicted infinite range
correlations are caused by an interaction between a light source and
a nearby scatterer\cite{Shapiro1999}. Since for a classical dipole
source this $C_0$ correlation is equivalent to fluctuations in the
LDOS\cite{Skipetrov2006, Carminati2010}, it is very likely that a
CPS will yield different results. We hope our work encourages the
use of more well-defined sources in theory and will help in choosing
the right type of source for the desired measurement.

\textit{Note - } While this manuscript was finalized, a theoretical
paper by Greffet et al. \cite{Greffet2010} was published where a
similar concept was developed emphasizing electronic circuit
analogies in the field of nanoantennas.

\begin{acknowledgments}
We thank Allard Mosk and Willem Vos for stimulating discussions.
Timmo van der Beek is acknowledged for help with the sample
preparation. This work is part of the research program of the
``Stichting voor Fundamenteel Onderzoek der Materie (FOM)'', which
is financially supported by the ``Nederlandse Organisatie voor
Wetenschappelijk Onderzoek (NWO)''.
\end{acknowledgments}

\bibliography{references}

\begin{thebibliography}{18}
\expandafter\ifx\csname natexlab\endcsname\relax\def\natexlab#1{#1}\fi
\expandafter\ifx\csname bibnamefont\endcsname\relax
  \def\bibnamefont#1{#1}\fi
\expandafter\ifx\csname bibfnamefont\endcsname\relax
  \def\bibfnamefont#1{#1}\fi
\expandafter\ifx\csname citenamefont\endcsname\relax
  \def\citenamefont#1{#1}\fi
\expandafter\ifx\csname url\endcsname\relax
  \def\url#1{\texttt{#1}}\fi
\expandafter\ifx\csname urlprefix\endcsname\relax\def\urlprefix{URL }\fi
\providecommand{\bibinfo}[2]{#2}
\providecommand{\eprint}[2][]{\url{#2}}

\bibitem[{\citenamefont{Wilk et~al.}(2007)\citenamefont{Wilk, Webster, Kuhn,
  and Rempe}}]{Wilk2007}
\bibinfo{author}{\bibfnamefont{T.}~\bibnamefont{Wilk}},
  \bibinfo{author}{\bibfnamefont{S.~C.} \bibnamefont{Webster}},
  \bibinfo{author}{\bibfnamefont{A.}~\bibnamefont{Kuhn}}, \bibnamefont{and}
  \bibinfo{author}{\bibfnamefont{G.}~\bibnamefont{Rempe}},
  \bibinfo{journal}{Science} \textbf{\bibinfo{volume}{317}},
  \bibinfo{pages}{488} (\bibinfo{year}{2007}).

\bibitem[{\citenamefont{Sapienza et~al.}(2010)\citenamefont{Sapienza,
  Thyrrestrup, Stobbe, Garcia, Smolka, and Lodahl}}]{Sapienza2010}
\bibinfo{author}{\bibfnamefont{L.}~\bibnamefont{Sapienza}},
  \bibinfo{author}{\bibfnamefont{H.}~\bibnamefont{Thyrrestrup}},
  \bibinfo{author}{\bibfnamefont{S.}~\bibnamefont{Stobbe}},
  \bibinfo{author}{\bibfnamefont{P.~D.} \bibnamefont{Garcia}},
  \bibinfo{author}{\bibfnamefont{S.}~\bibnamefont{Smolka}}, \bibnamefont{and}
  \bibinfo{author}{\bibfnamefont{P.}~\bibnamefont{Lodahl}},
  \bibinfo{journal}{Science} \textbf{\bibinfo{volume}{327}},
  \bibinfo{pages}{1352} (\bibinfo{year}{2010}).

\bibitem[{\citenamefont{Koenderink et~al.}(2002)\citenamefont{Koenderink,
  Bechger, Schriemer, Lagendijk, and Vos}}]{Koenderink2002}
\bibinfo{author}{\bibfnamefont{A.~F.} \bibnamefont{Koenderink}},
  \bibinfo{author}{\bibfnamefont{L.}~\bibnamefont{Bechger}},
  \bibinfo{author}{\bibfnamefont{H.~P.} \bibnamefont{Schriemer}},
  \bibinfo{author}{\bibfnamefont{A.}~\bibnamefont{Lagendijk}},
  \bibnamefont{and} \bibinfo{author}{\bibfnamefont{W.~L.} \bibnamefont{Vos}},
  \bibinfo{journal}{Phys. Rev. Lett.} \textbf{\bibinfo{volume}{88}},
  \bibinfo{pages}{143903} (\bibinfo{year}{2002}).

\bibitem[{\citenamefont{Farahani et~al.}(2005)\citenamefont{Farahani, Pohl,
  Eisler, and Hecht}}]{Farahani2005}
\bibinfo{author}{\bibfnamefont{J.~N.} \bibnamefont{Farahani}},
  \bibinfo{author}{\bibfnamefont{D.~W.} \bibnamefont{Pohl}},
  \bibinfo{author}{\bibfnamefont{H.~J.} \bibnamefont{Eisler}},
  \bibnamefont{and} \bibinfo{author}{\bibfnamefont{B.}~\bibnamefont{Hecht}},
  \bibinfo{journal}{Phys. Rev. Lett.} \textbf{\bibinfo{volume}{95}},
  \bibinfo{pages}{017402} (\bibinfo{year}{2005}).

\bibitem[{\citenamefont{T\"ureci et~al.}(2008)\citenamefont{T\"ureci, Ge,
  Rotter, and Stone}}]{Tureci2008}
\bibinfo{author}{\bibfnamefont{H.~E.} \bibnamefont{T\"ureci}},
  \bibinfo{author}{\bibfnamefont{L.}~\bibnamefont{Ge}},
  \bibinfo{author}{\bibfnamefont{S.}~\bibnamefont{Rotter}}, \bibnamefont{and}
  \bibinfo{author}{\bibfnamefont{A.~D.} \bibnamefont{Stone}},
  \bibinfo{journal}{Science} \textbf{\bibinfo{volume}{320}},
  \bibinfo{pages}{643} (\bibinfo{year}{2008}).

\bibitem[{\citenamefont{Tsien}(1998)}]{Tsien1998}
\bibinfo{author}{\bibfnamefont{R.~Y.} \bibnamefont{Tsien}},
  \bibinfo{journal}{Annu. Rev. Biochem.} \textbf{\bibinfo{volume}{67}},
  \bibinfo{pages}{509} (\bibinfo{year}{1998}).

\bibitem[{\citenamefont{Westphal and Hell}(2005)}]{Hell2005}
\bibinfo{author}{\bibfnamefont{V.}~\bibnamefont{Westphal}} \bibnamefont{and}
  \bibinfo{author}{\bibfnamefont{S.~W.} \bibnamefont{Hell}},
  \bibinfo{journal}{Phys. Rev. Lett.} \textbf{\bibinfo{volume}{94}},
  \bibinfo{pages}{143903} (\bibinfo{year}{2005}).

\bibitem[{\citenamefont{Min et~al.}(2009)\citenamefont{Min, Lu, Chong, Roy,
  Holtom, and Xie}}]{Min2009}
\bibinfo{author}{\bibfnamefont{W.}~\bibnamefont{Min}},
  \bibinfo{author}{\bibfnamefont{S.}~\bibnamefont{Lu}},
  \bibinfo{author}{\bibfnamefont{S.}~\bibnamefont{Chong}},
  \bibinfo{author}{\bibfnamefont{R.}~\bibnamefont{Roy}},
  \bibinfo{author}{\bibfnamefont{G.}~\bibnamefont{Holtom}}, \bibnamefont{and}
  \bibinfo{author}{\bibfnamefont{X.~S.} \bibnamefont{Xie}},
  \bibinfo{journal}{Nature} \textbf{\bibinfo{volume}{461}},
  \bibinfo{pages}{1105} (\bibinfo{year}{2009}).

\bibitem[{\citenamefont{Horowitz and Hill}(1989)}]{Horowitz}
\bibinfo{author}{\bibfnamefont{P.}~\bibnamefont{Horowitz}} \bibnamefont{and}
  \bibinfo{author}{\bibfnamefont{W.}~\bibnamefont{Hill}},
  \emph{\bibinfo{title}{The Art of Electronics}} (\bibinfo{publisher}{Cambridge
  University Press}, \bibinfo{year}{1989}).

\bibitem[{\citenamefont{Wiersma}(2008)}]{Wiersma2008}
\bibinfo{author}{\bibfnamefont{D.~S.} \bibnamefont{Wiersma}},
  \bibinfo{journal}{Nature Physics} \textbf{\bibinfo{volume}{4}},
  \bibinfo{pages}{359} (\bibinfo{year}{2008}).

\bibitem[{\citenamefont{Siegman}(1986)}]{Siegman}
\bibinfo{author}{\bibfnamefont{A.}~\bibnamefont{Siegman}},
  \emph{\bibinfo{title}{Lasers}} (\bibinfo{publisher}{University Science
  Books}, \bibinfo{year}{1986}).

\bibitem[{\citenamefont{Isak and Eyring}(1992)}]{Isak1992}
\bibinfo{author}{\bibfnamefont{S.~J.} \bibnamefont{Isak}} \bibnamefont{and}
  \bibinfo{author}{\bibfnamefont{E.~M.} \bibnamefont{Eyring}},
  \bibinfo{journal}{J. Phys. Chem.} \textbf{\bibinfo{volume}{96}},
  \bibinfo{pages}{1738} (\bibinfo{year}{1992}).

\bibitem[{\citenamefont{van Soest and Lagendijk}(2002)}]{GvS2002}
\bibinfo{author}{\bibfnamefont{G.}~\bibnamefont{van Soest}} \bibnamefont{and}
  \bibinfo{author}{\bibfnamefont{A.}~\bibnamefont{Lagendijk}},
  \bibinfo{journal}{Phys. Rev. E} \textbf{\bibinfo{volume}{65}},
  \bibinfo{pages}{047601} (\bibinfo{year}{2002}).

\bibitem[{\citenamefont{Cao et~al.}(1999)\citenamefont{Cao, Zhao, Ho, Seelig,
  Wang, and Chang}}]{Cao1999}
\bibinfo{author}{\bibfnamefont{H.}~\bibnamefont{Cao}},
  \bibinfo{author}{\bibfnamefont{Y.~G.} \bibnamefont{Zhao}},
  \bibinfo{author}{\bibfnamefont{S.~T.} \bibnamefont{Ho}},
  \bibinfo{author}{\bibfnamefont{E.~W.} \bibnamefont{Seelig}},
  \bibinfo{author}{\bibfnamefont{Q.~H.} \bibnamefont{Wang}}, \bibnamefont{and}
  \bibinfo{author}{\bibfnamefont{R.~P.~H.} \bibnamefont{Chang}},
  \bibinfo{journal}{Phys. Rev. Lett.} \textbf{\bibinfo{volume}{82}},
  \bibinfo{pages}{2278} (\bibinfo{year}{1999}).

\bibitem[{\citenamefont{Shapiro}(1999)}]{Shapiro1999}
\bibinfo{author}{\bibfnamefont{B.}~\bibnamefont{Shapiro}},
  \bibinfo{journal}{Phys. Rev. Lett.} \textbf{\bibinfo{volume}{83}},
  \bibinfo{pages}{4733} (\bibinfo{year}{1999}).

\bibitem[{\citenamefont{van Tiggelen and Skipetrov}(2006)}]{Skipetrov2006}
\bibinfo{author}{\bibfnamefont{B.~A.} \bibnamefont{van Tiggelen}}
  \bibnamefont{and} \bibinfo{author}{\bibfnamefont{S.~E.}
  \bibnamefont{Skipetrov}}, \bibinfo{journal}{Phys. Rev. E}
  \textbf{\bibinfo{volume}{73}}, \bibinfo{pages}{045601}
  (\bibinfo{year}{2006}).

\bibitem[{\citenamefont{Caz\'e et~al.}(2010)\citenamefont{Caz\'e, Pierrat, and
  Carminati}}]{Carminati2010}
\bibinfo{author}{\bibfnamefont{A.}~\bibnamefont{Caz\'e}},
  \bibinfo{author}{\bibfnamefont{R.}~\bibnamefont{Pierrat}}, \bibnamefont{and}
  \bibinfo{author}{\bibfnamefont{R.}~\bibnamefont{Carminati}},
  \bibinfo{journal}{Phys. Rev. A} \textbf{\bibinfo{volume}{82}},
  \bibinfo{pages}{043823} (\bibinfo{year}{2010}).

\bibitem[{\citenamefont{Greffet et~al.}(2010)\citenamefont{Greffet, Laroche,
  and Marquier}}]{Greffet2010}
\bibinfo{author}{\bibfnamefont{J.~J.} \bibnamefont{Greffet}},
  \bibinfo{author}{\bibfnamefont{M.}~\bibnamefont{Laroche}}, \bibnamefont{and}
  \bibinfo{author}{\bibfnamefont{F.}~\bibnamefont{Marquier}},
  \bibinfo{journal}{Phys. Rev. Lett.} \textbf{\bibinfo{volume}{105}},
  \bibinfo{pages}{117701} (\bibinfo{year}{2010}).

\end{thebibliography}
\pagebreak


\section*{Supplementary material}

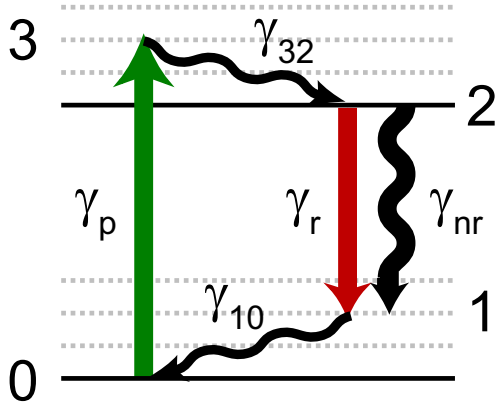

FIG. 1: (a) Jablonski diagram of a four level system. Wiggly arrows represent nonradiative transitions, straight arrwow represent radiative transitions.

## I. DERIVATION OF GENERALIZED EXPRESSION FOR A SOURCE

Our goal is to generalize the axiomatic expression for a Constant Amplitude Source (CAS),

$$j(\mathbf{r},t) = j_0 \delta(\mathbf{r}-\mathbf{r_0})\exp(-i\omega t) + \text{c.c.}, \tag{1}$$

in such a way that the new expression fully incorporates both Constant Power Sources (CPS) and CAS. For illustration purposes we study a four-level system, but our approach is general and is applicable to other optical transitions as well. We start by writing down four equations for the population of the different energy levels, that are shown in a Jablonski diagram in Fig. 1(a). Molecules are excited from the ground state (level 0) to the excited state (level 3) by a pump rate $\gamma_\text{p}$. Rapid nonradiative transitions let molecules decay from this excited state 3 to a lower-lying state 2. Molecules can then decay either radiatively by spontaneous emission with rate $\gamma_\text{r}$ or nonradiatively with rate $\gamma_\text{nr}$ to a vibrational sublevel of the ground state (level 1). The radiative transition from level 2 to level 1 constitutes our light source. The rate equations for the four levels read

$$\frac{dN_0}{dt} = -\gamma_\text{p} N_0 + \gamma_{10} N_1 \tag{2}$$

$$\frac{dN_3}{dt} = \gamma_\text{p} N_0 - \gamma_{32} N_3, \tag{3}$$

$$\frac{dN_2}{dt} = \gamma_{32} N_3 - (\gamma_\text{r} + \gamma_\text{nr}) N_2, \tag{4}$$

$$\frac{dN_1}{dt} = (\gamma_\text{r} + \gamma_\text{nr}) N_2 - \gamma_{10} N_1. \tag{5}$$

We are now particularly interested in finding the stationary rate of photon production, $\gamma_r N_2$, expressed in terms of the pump rate and the radiative and non-radiative decay rate. Using $\gamma_{10} N_1 = (\gamma_r + \gamma_{nr}) N_2$ from Eq. (5) in Eq. (2) gives

$$N_2 = N_0 \frac{\gamma_p}{\gamma_r + \gamma_{nr}}, \tag{6}$$

and hence for the photon production rate

$$\gamma_r N_2 = N_0 \gamma_r \frac{\gamma_p}{\gamma_r + \gamma_{nr}}, \tag{7}$$

$$= \gamma_r \frac{\gamma_e}{\gamma_r + \gamma_{nr}}. \tag{8}$$

In general the depopulation of the ground level, $N_0$, in a four-level system is negligible and the factor $\gamma_p N_0$ can be taken as the constant effective excitation rate: $\gamma_e \equiv \gamma_p N_0$.

In the case of a CAS the emitted power, which is proportional to the photon production rate, is given by

$$P_{\text{src}}^{\text{CAS}}/P_0 = -\frac{4\pi c}{\omega} \text{Im} G_\omega(\mathbf{r_0}, \mathbf{r_0}) \equiv \frac{4\pi^2 c^3}{\omega^2} \text{LDOS}(\mathbf{r_0}, \omega) = \frac{4\pi^2 c^3}{\omega^2} \frac{\gamma_r}{A}, \tag{9}$$

while we just found that based on a straightforward steady-state analysis of a four-level system the power is proportional to $\gamma_r \frac{\gamma_e}{\gamma_r + \gamma_{nr}}$. Clearly the expression for a source from which Eq. (9) is deduced is not complete. Hence we adjust the source term such that the resulting power becomes proportional to $\gamma_r \frac{\gamma_e}{\gamma_r + \gamma_{nr}}$, that is

$$P_{\text{src}}/P_0 = \frac{\gamma_r}{\gamma_r + \gamma_{nr}} \Big/ \frac{\gamma_r^{(0)}}{\gamma_r^{(0)} + \gamma_{nr}} = \frac{4\pi^2 c^3}{\omega^2} \text{LDOS}(\mathbf{r_0}, \omega) \frac{\gamma_r^{(0)} + \gamma_{nr}}{\gamma_r + \gamma_{nr}}. \tag{10}$$

this equation implies the original expression (1) for a CAS source needs to be adjusted to

$$j(\mathbf{r}, t) = \sqrt{\frac{\gamma_e}{\gamma_r + \gamma_{nr}}} \delta(\mathbf{r} - \mathbf{r_0}) \exp(-i\omega t) + \text{c.c.}. \tag{11}$$

This is our generalized expression for a light source. In the case of a CPS ($\gamma_{nr} = 0$), the output power is independent of the environment, whereas in the case of a CAS ($\gamma_{nr} \gg \gamma_r$) the output power depends on the radiative decay rate and thus the local environment of the emitter. We note that in our derivation we have assumed a constant (non-photonic) excitation rate of the four-level system, our analysis needs to be expanded when the system is pumped into saturation. Since in the case of saturation a change in the total decay rate will also change the excitation rate.

## II. STIMULATED EMISSION AND THE LDOS

In the analysis given in Sec. I stimulated emission was not considered: the light source was formed by spontaneous radiative decay from level 2 to level 1. However in a random laser above threshold stimulated emission is the main mechanism of radiation. In this section, we show how stimulated emission can be incorporated into our analysis of CPS and CAS. In fact the laser rate equations describing the number of molecules in the upper laser level is very similar to Eq. (4) if we consider $R = \gamma_p N_0 = \gamma_e$ and $\gamma_{tot} = \gamma_r + \gamma_{nr}$:

$$\frac{dN}{dt} = R - N\gamma_{tot} - \beta q N \gamma_r, \qquad (12)$$

$$= \gamma_e - (\gamma_r + \gamma_{nr} + \gamma_r \beta q)N. \qquad (13)$$

The only extra term appearing in the above equation is due to stimulated emission $\beta \gamma_r N q$. In this equation it is assumed that reabsorption from level 1 to level 2 can be ignored. This assumption is realistic since the vibrational relaxation from level 1 to level 0 is often very fast. In a laser, photons are confined in a cavity configuration and the created photons give feed-back onto the population rate equations. For didactic purposes, we here neglect the feedback from these created photons on the source and assume most photons that induce stimulated emission originate from elsewhere. Stimulated emission then appears in Eq. (4) and (5) as an extra term $W_{21}N_2 = B_{21}\rho N_2$, where $W_{21}$ is the stimulated emission rate, $B_{21}$ is the Einstein coefficient for stimulated emission, and $\rho$ is the energy density of the impinging photons[1]. The photon production rate must now take into account both stimulated emission and spontaneous emission, we find

$$(\gamma_r + B_{21}\rho)N_2 = (\gamma_r + B_{21}\rho)\frac{\gamma_e}{\gamma_r + B_{21}\rho + \gamma_{nr}}. \qquad (14)$$

From this equation we can now fully understand the effect of stimulated emission on the light source. In the case of a CPS, $\gamma_{nr} = 0$ and the photon production rate is only dependent on the pump rate. In the case of a CAS, $\gamma_{nr} \gg \gamma_r$, stimulated emission increases the total number of photons that is produced. In that sense stimulated emission and LDOS fluctuations lead to similar effects on the emission of photons by a source.

In addition, we note that for $\rho \to \infty$ a CAS is turned into a CPS showing that stimulated emission can be a convenient tool for engineering the properties of a CAS light source.

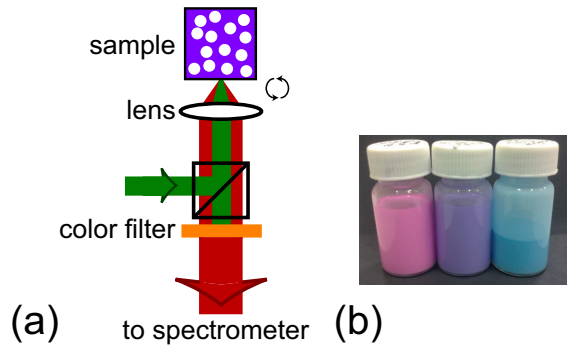

FIG. 2: (a) Experimental apparatus to study light sources in random lasers. Green: pump beam. Red: emission light. (b) Photograph of the three random laser suspensions. From left to right: Rhodamine 640 P, Cresyl Violet, and Nile Blue.

## III. EXPERIMENTAL METHODS

Three molecular light sources were studied in a random laser configuration by suspending titania particles (R900 DuPont, volume fraction 1%) into three different 1 mM solutions of organic dyes in methanol. The three dye solutions acted as gain media and were chosen based on their quantum yields ($\phi$) reported in literature [2, 3]: Rhodamine 640 P ($\phi = 1$), Cresyl Violet ($\phi = 0.54$), and Nile Blue ($\phi = 0.27$). A photograph of the samples is shown in Fig 2(b). To prevent aggregation and sedimentation of titania particles all samples were treated in an ultrasonic bath before and spinned during measurement, and a small amount of $CaCl_2$ (0.06 g/L) was added to the Nile Blue sample. Quartz cuvettes were used as experimental cells (Hellma, inner dimensions 10×10×45 mm, wall thickness 1.25 mm).

The experimental apparatus is shown in Fig 2(a). Excitation light generated by an optical parametric oscillator (Opolette, 20 Hz, 5 ns) was focussed onto the samples by an aspherical lens (F/#=1.5). The same lens collected the emission which was then spectrally analyzed using a spectrograph (Oriel MS-257) connected to an EMCCD camera (Hamamatsu, C-9100).

[1] O. Svelto, *Principles of Lasers* (Plenum Press, 1998).

[2] A. Siegman, *Lasers* (University Science Books, 1986).

[3] S. J. Isak and E. M. Eyring, J. Phys. Chem. **96**, 1738 (1992).



\end{document}